\documentclass[12pt,aps,superscriptaddress,longbibliography]{revtex4-1}
\usepackage{amsmath}
\usepackage{amsfonts}
\usepackage{amssymb}
\usepackage{graphicx}
\usepackage{latexsym}
\usepackage{braket}
\usepackage{subfig}
\usepackage{pifont}
\usepackage{color}
\usepackage[margin=1in]{geometry}
\usepackage{hyperref}
\usepackage[compact]{titlesec}         
\titlespacing{\section}{0pt}{0pt}{0pt} 
\AtBeginDocument{
  \setlength\abovedisplayskip{0pt}
  \setlength\belowdisplayskip{0pt}}
\makeatletter
\let\cat@comma@active\@empty
\makeatother
\begin{document}
\title{Emergence of a new symmetry class for  Bogoliubov-de Gennes (BdG) Hamiltonians:  Expanding ten-fold  symmetry classes}
\author{Ranjith Kumar R}
\affiliation{Poornaprajna Institute of Scientific Research, 4, 16th Cross,
	Sadashivnagar, \\Bengaluru - 5600-80, India.}
\affiliation{Manipal Academy of Higher Education, Madhava Nagar, \\Manipal - 576104, India.}
\author{Sujit Sarkar}
\affiliation{Poornaprajna Institute of Scientific Research, 4, 16th Cross,
	Sadashivnagar, \\Bengaluru - 5600-80, India.}
	\date{\today}

\begin{abstract}
\noindent Symmetry plays an important role in the topological states of quantum matter. Cartan found three symmetry classes, Altland and Zirnbauer extended it to the ten-fold symmetry class for mesoscopic systems. We observe emergence of a new symmetry class for BdG Hamiltonians. This new symmetry class appears for the BdG Hamiltonian in spinful background. We also observe the interesting feature that there is no evidence of Kramers's degeneracy for this BdG model Hamiltonian either in spinless or in spinful background.  We call the extra symmetry class $X$, i.e, expanding the ten-fold symmetry classes for topological states of quantum matter. These BdG Hamiltonians show many new, interesting and insightful results.
\end{abstract}

\maketitle
\section{Introduction}
A symmetry is a transformation which leaves the physical system invariant. These transformations include translation, reflection, rotation, scaling, etc. One of the most important implications of symmetry in physics is the existence of conservation laws. For every global continuous symmetry there exists an associated conserved quantity \cite{Noether:1918zz}. In quantum mechanics symmetry transformation can be represented on the Hilbert space of physical states by an operator that is either linear and unitary or anti-linear and anti-unitary \cite{bargmann1964note}. Any symmetry operator acts on these states and transforms them to new states. These symmetry operators can be classified as continuous (rotation, translation) and discrete (parity, lattice translations, time reversal). Continuous symmetry transformations give rise to conservation of probabilities and discrete symmetry transformations give rise to the quantum numbers. Another important implication of symmetry in quantum mechanics is the symmetry on exchanging identical particles \cite{QFT:gifted}. \\
To classify the different states of matter, Landau, in the year 1950 developed a theory, called Landau's theory of spontaneous symmetry breaking, based on the idea of order parameter \cite{Ginzburg:1950sr}. According to Landau phase transition always occurs with spontaneous breaking of symmetry. In other words, if two phases cannot be continuously connected without crossing a phase boundary, they must have different symmetries. \\
Landau's theory of spontaneous symmetry breaking was not the complete picture of classifying the different states of matter. In fact, at any finite temperature, symmetry of many-body systems cannot be spontaneously broken in dimension $d\leq2$ \cite{mermin1966absence,PhysRev.158.383}. After the discovery of quantum Hall effect \cite{QHE-von-klitzing,PhysRevLett.48.1559}, a new way of classifying the different states of matter based on their topological properties has been developed \cite{simons1997phase,PhysRevLett.95.146802,qi2011topological,wu2006helical}. Even in this classification system symmetry plays an important role \cite{slager2013space,Classifi-symmetry,PhysRevX.7.041069}. \\
\textbf{Relevance and motivation of this study :} In theoretical physics different emergent particles appear as a Bogoliubov quasiparticles \cite{stanescu2016}. Here we state it very briefly. This Bogoliubov quasiparticles appear also as Majorana neutrinos of particle physics \cite{majorana-nutrino}. The other realization of Bogoliubov quasiparticles is in quantum many-body physics, quantum optics and specially in topological state of quantum many body systems \cite{sarkar2nature}. The Bogoliubov quasiparticle appears in the solution of the Bogoliubov-de Gennes (BdG) Hamiltonian.\\

\noindent Among the vast variety of topological phases one can identify
an important class called symmetry protected topological (SPT) phase, where two quantum
states have distinct topological properties protected by certain symmetry \cite{stanescu2016, Fu-SPT}. An interesting candidate for topological quantum states,
topological superconductors (TSCs) have fully gapped bulk but gapless conducting surface
states.
The surface states in TSCs are Bogoliubov quasiparticles with equal parts of electron and hole excitations \cite{stanescu2016}. Because of the particle-hole symmetry of the superconductors these zero energy Bogoliubov quasiparticles can be realized as Majorana zero modes \cite{volovik1999fermion,senthil2000quasiparticle,kitaev2001}. There are some promising proposals for this purpose \cite{lutchyn2010majorana,nadj2014observation,murakawa2009new}. Another example of SPT states are topological nodal systems, which are protected by the non-spatial symmetries and spatial lattice symmetries. Nodal non-centrosymmetric superconductors are one of the realizations of materials of this kind \cite{brydon2011topologically,schnyder2011topological}. The Bogoliubov quasiparticle appears in a different context in the solution of the Bogoliubov-de Gennes (BdG) Hamiltonian.\\
Bogoliubov transformation of superconductor (s-wave or p-wave) gives the energy spectrum of quasiparticle excitation. These Bogoliubov quasiparticles are combinations of particles and holes. The mean-field superconductor Hamiltonian defined in terms of first quantized Hamiltonian of single particle excitations helps in realization of Bogoliubov quasiparticles \cite{stanescu2016}. The first quantized Hamiltonian or Bogoliubov-de Gennes Hamiltonian for 1D and 2D p-wave superconductors, helps in the realization of Majorana and chiral Majorana zero modes respectively \cite{wilczek2009majorana}. The Majorana neutrinos of particle physics can also be found as Bogoliubov quasiparticles \cite{majorana-nutrino}. Thus it is clear from this discussion, the importance of Bogoliubov quasiparticles as an emergent particles in different branches of theoretical physics.\\
In this theoretical study, we have introduced a set of BdG Hamiltonians. We study the symmetry classes of these BdG
Hamiltonians for both spinless and spinful fermionic background. This study results in a new symmetry class in the spinful background of the model Hamiltonians. This new symmetry class extends the existing ten-fold symmetry class of topological state of matter. We also search properties of Kramer's degeneracy for this BdG Hamiltonian in both spinful and spinless fermionic background.\\ 
\textbf{Plan of this research article :} We present the basic aspects of symmetries for fermionic systems in section II. In section III, we present the model Hamiltonians. In section IV we present the results and discussion for symmetries of model Hamiltonians in spinless background. The results and discussion for symmetries of model Hamiltonians in spinful background are in section V.  
\section{Symmetries in fermionic systems}
The BdG Hamiltonians we consider here are for the fermionic system. Therefore we present briefly what kind of symmetries are present for the fermionic system.
In fermionic systems one can write the second quantized Hamiltonian for non-superconducting systems using the fermionic annihilation or creation operators, which obey the canonical anti-commutation relation
$ \left\lbrace \psi_i, \psi_j^{\dagger}\right\rbrace = \delta_{ij}$ as
\begin{equation}
\mathcal{H}=\psi^{\dagger}H\psi,
\end{equation}
where $H$ is the first quantized $N\times N$ matrix. In quantum mechanics a symmetry operation is represented by the operator acting on the Hilbert space. Thus the symmetry transformation on the fermion operator can be represented as
\begin{equation}
\mathcal{O}\psi_i\mathcal{O}^{-1}= O_i^j \psi_j.
\end{equation} 
If the system is invariant under the transformation $\mathcal{O}$ then the canonical anti-commutation relation and the Hamiltonian $H$ should be preserved, i.e, $\left\lbrace \psi_i, \psi_j^{\dagger}\right\rbrace = \mathcal{O} \left\lbrace \psi_i, \psi_j^{\dagger}\right\rbrace \mathcal{O}^{-1}$ and \; $\mathcal{O} H \mathcal{O}^{-1}=H$. This symmetry operation $\mathcal{O}$ can be either spatial or non-spatial. When  $\mathcal{O}$ is spatial i.e, $\mathcal{O}=\Pi_i \mathcal{O}_i$, it acts on each lattice site separately, otherwise it is called non-spatial or         internal symmetry transformation. \\
The second quantized Hamiltonian for a superconducting system can be described in terms of single-particle s-wave pairing mean field Hamiltonian or Bogoliubov-de Gennes (BdG) Hamiltonian, 
\begin{equation}
\mathcal{H}=\frac{1}{2} \Psi^{\dagger}H_{BdG}\Psi,
\end{equation}
where $\Psi^{\dagger}$ and $\Psi$ are the four-component Nambu spinors instead of complex fermion operators. $H_{BdG}$ is the first quantized Bogoliubov-de Gennes Hamiltonian which governs the dynamics of Bogoliubov quasiparticle and is given by
\begin{equation}
H_{BdG}= \left( \begin{matrix}
H_0 && -i\sigma_y\Delta\\
i\sigma_y\Delta^{*} && -H_0^T\\
\end{matrix}\right) ,
\end{equation}
where $H_0$ and $\Delta$ are kinetic and pair potential terms respectively and $\sigma_y$ is one of the Pauli matrices.
This BdG theory has an intrinsic particle-hole constraint. Below we discuss some of the non-spatial symmetry transformations which are important in the understanding of symmetry protected topological states.\\
Different SPT states can be well understood with the local (gauge) non-spatial symmetries such as time-reversal (TR), particle-hole (PH) and chiral. In general non-interacting Hamiltonians can be classified in terms of symmetries into ten different symmetry classes \cite{Altland1997}. A particular symmetry class of a Hamiltonian is determined by its invariance under time-reversal, particle-hole and chiral symmetries. This is because TR and PH are the anti-unitary symmetries which cannot be removed by block-diagonalization of the Hamiltonian. Chiral symmetry is the unitary symmetry anti-commuting with the Hamiltonian and it is defined as the product of TR and PH symmetries. Thus the study of TR, PH and chiral symmetries reveals important toplogical characters of SPT states. 
\subsection*{Time-reversal symmetry}
Time-reversal symmetry is an anti-unitary transformation which reverses the arrow of time. $\mathcal{T} : t\longrightarrow-t$, where $\mathcal{T}$ is the time-reversal operator. If a system with Hamiltonian $H$ is invariant with the reversal of time then we have the following commutation relation,
\begin{equation}
[\mathcal{T},H]=0, \;\;\;\; i.e. \;\; \mathcal{T}H\mathcal{T}^{-1}=H.
\end{equation} 
The time-reversal operator does not affect the position but it reverses the sign of momentum and also it acts as the complex conjugate operator($K$).
\begin{equation}
\mathcal{T}x\mathcal{T}^{-1}=x , \;\;\; \mathcal{T}k\mathcal{T}^{-1}=-k,\;\;\; \mathcal{T}i\mathcal{T}^{-1}=-i .
\end{equation}
Since $\mathcal{T}$ is an anti-unitary operator we can represent it as a product of unitary ($U$) and complex conjugate operators, i.e. $\mathcal{T}=UK$.  For spinless particles we can represent time-reversal operator as the complex conjugate operator and we have $\mathcal{T}^2=1$,
\begin{equation}
\mathcal{T}=K .
\end{equation}
For spin-$\frac{1}{2}$ particles we have $\mathcal{T}^2=-1$ and the operator is represented as
\begin{equation}
\mathcal{T}=i\sigma_y K.
\end{equation}
The square of the time-reversal operator equals negative of identity which yields to Kramer's degeneracy. According to that every state is doubly degenerate and one state is time-reversal of another. Thus the system becomes time-reversal invariant \cite{bernevig-topo-insu,Classifi-symmetry}.
\subsection*{Particle-hole symmetry}
Particle-hole (PH) symmetry is a transformation between electrons and holes within certain energy range. If a Hamiltonian has PH symmetry then we have the following anti-commutation relation \cite{stanescu2016,Kotetes2014,Classifi-symmetry},
\begin{equation}
\left\lbrace \mathcal{C},H\right\rbrace =0 \;\;\;\; i.e, \;\; \mathcal{C}H\mathcal{C}^{-1}=-H.
\end{equation}
The PH operator is anti-unitary operator with $\mathcal{C}^2=1$  and it is represented by
\begin{equation}
\mathcal{C}=\sigma_x K.
\end{equation} 
Systems with $SU(2)$ spin symmetry will have PH symmetry operator as
\begin{equation}
\mathcal{C}=i\sigma_y K.
\end{equation}
In this case we have the square of the PH symmetry as $\mathcal{C}^2=-1$. In the presence of PH symmetry each eigen-function $\Psi$ with $E>0$ has its particle-hole reversed partner, $\mathcal{C}\Psi$ with $E<0$. The PH symmetry is an intrinsic property of the mean field theory of superconductivity. 
\subsection*{Chiral symmetry}
The product of time-reversal operator ($\mathcal{T}$) and particle-hole operator ($\mathcal{C}$) leads to a third kind of symmetry called chiral symmetry($\mathcal{S}$) or sub-latticle symmetry. The symmetry operator is given by
\begin{equation}
\mathcal{S}=\sigma_x.
\end{equation} 
It is a unitary symmetry tranformation which anticommutes with the Hamiltonian with $\mathcal{S}^2=1$,
\begin{equation}
\left\lbrace \mathcal{S},H \right\rbrace =-H, \;\;\;\; i.e. \;\; \mathcal{S}H\mathcal{S}^{-1}=-H.
\end{equation} 
Chiral symmetry results in the symmetric spectrum. From Table 1 one can observe that presence of either TR or PH results in absence of chiral symmetry. Also, absence of both TR and PH symmetries results in either presence or absence of chiral symmetry \cite{stanescu2016}.\\
\begin{table}
\begin{center}
\includegraphics[scale=0.2]{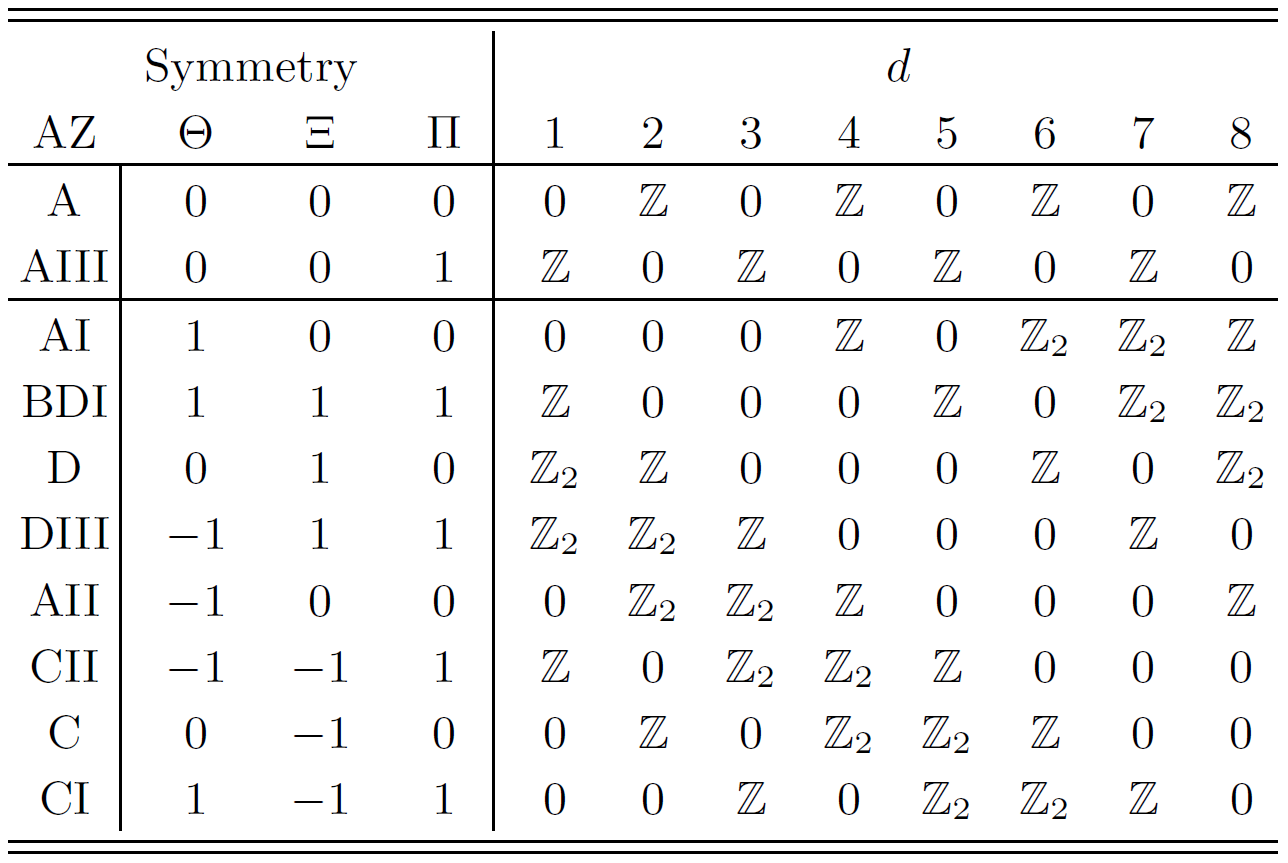}
\end{center}
\caption{Periodic table of topological insulators and topological superconductors. Here $\Theta$ is time-reversal, $\Xi$ is particle-hole and $\Pi$ is chiral symmetry operators.}
\label{periodictable}
\end{table}
Based on the behaviour of Hamiltonian with the TR, PH and chiral symmetries, it is classified in ten symmetry classes \cite{Altland1997}. Table \ref{periodictable} shows the ten symmetry classes of non-interacting Hamiltonian. One can identify for given spatial dimension $d$ of a system and given set of symmetries which symmetry class it belongs to and its topological properties. $\mathbb{Z}$ and $\mathbb{Z}_2$ are the topological invariant numbers which take integer values ($\mathbb{Z}=0,\pm1,\pm2,...$ and $\mathbb{Z}_2=\pm1$), and represent the topological distinct phases within a given symmetry class.\\
A BdG Hamiltonian in general possesses different symmetry classes based on different constraints. In general BdG Hamiltonians can fall under one among the following symmetry classes.\\
\textbf{Class D:} The BdG Hamiltonians are described in terms of Nambu spinors,
\begin{equation}
\Psi= \left( \begin{matrix}
\psi_{1},...,\psi_{N},\psi^{\dagger}_{1},...,\psi^{\dagger}_{N}
\end{matrix}\right)^T, \;\;\;\;\;\;\;   \Psi^{\dagger}=\left( \begin{matrix}
 \psi^{\dagger}_{1},...,\psi^{\dagger}_{N}, \psi_{1},...,\psi_{N},\\
\end{matrix}\right).
\end{equation}
where $\Psi$ and $\Psi^{\dagger}$ obey the the canonical anti-commutation relation $\left\lbrace \Psi_i, \Psi^{\dagger}_j\right\rbrace=\delta_{ij} $. Thus Nambu spinors $\Psi$ and $\Psi^{\dagger}$ are not independent of each other. This dependency places a constraint on the Hamiltonian of the system as
\begin{equation}
\sigma_x H^T \sigma_x=-H,\label{D}
\end{equation}
where $\sigma_x$ is one of the Pauli matrices. This constraint is a built-in feature of BdG Hamiltonians that originates from Fermi statistics. Thus single-particle BdG Hamiltonians are characterized by the PH constraint.\\
\textbf{Class DIII:} The set of BdG Hamiltonians with TR symmetry forms class DIII. The TRS condition implies
\begin{equation}
\sigma_y H^* \sigma_y=H, \label{DIII}
\end{equation}
where $\sigma_y$ is a Pauli spin matrix. Thus the class DIII Hamiltonians have both the constraints (eq.\ref{DIII} and eq.\ref{D}). The combined PH and TR symmetries give rise to chiral symmetry of the form $ \sigma_x\sigma_y H \sigma_x\sigma_y=H$. When TR symmetry operator, $\mathcal{T}^2=+1$, it represents BDI symmetry class.\\
\textbf{Classes A and AIII:} The BdG Hamiltonians with a $U(1)$ spin-rotation symmetry around the $S_z$ axis in spin space have no constraints and
\begin{equation}
\Psi^{\dagger}=\left( \begin{matrix}
\psi_{I \uparrow}^{\dagger} && \psi_{I \downarrow} 
\end{matrix}\right) , \;\;\;\;\;\;\; \Psi=\left( \begin{matrix}
\psi_{I \uparrow}\\
\psi_{I \downarrow}^{\dagger}
\end{matrix}\right).
\end{equation}
Unlike in class D, here $\Psi$ and $\Psi^{\dagger}$ are independent of each other. Thus the Hamiltonian H without any constraint forms class A. With the independent nature of $\Psi$ and $\Psi^{\dagger}$ one can write $\Psi^{\dagger}_{\downarrow,\uparrow}\rightarrow\Psi_{\downarrow,\uparrow}$. Also by imposing TRS one can obtain
\begin{equation}
\mathcal{T}\Psi\mathcal{T}^{-1}=\left( \begin{matrix}
\Psi_{\downarrow}\\\
-\Psi^{\dagger}_{\uparrow}\\
\end{matrix}\right) = i\sigma_y(\Psi^{\dagger})^T=\Psi^c.
\end{equation}
Thus by imposing $\Psi^{\dagger}_{\downarrow,\uparrow}\rightarrow\Psi_{\downarrow,\uparrow}$ and TR symmetry we have the effect which can be interpreted as the combination of TRS and PHS. In other words the system has  effective chiral symmetry. Thus the BdG Hamiltonians with this chiral symmetry form class AIII.\\
\textbf{Classes C and CI:} The BdG Hamiltonians with $SU(2)$ spin-rotation symmetry form the groups C and CI. The spin rotataion $\mathcal{O}^{\phi}$ around the $S_x$ transforms $\Psi$ into
\begin{equation}
\mathcal{O}^{\phi}_{S_x} \Psi \mathcal{O}^{-\phi}_{S_x}=\cos(\frac{\phi}{2})\Psi - i\sin(\frac{\phi}{2})\Psi^c.
\end{equation}
Also, spin-rotataion $\mathcal{O}^{\phi}$ around the $S_y$ transforms $\Psi$ into
\begin{equation}
\mathcal{O}^{\phi}_{S_y} \Psi \mathcal{O}^{-\phi}_{S_y}=\cos(\frac{\phi}{2})\Psi - \sin(\frac{\phi}{2})\Psi^c.
\end{equation}
Spin rotation by $\pi$ around $S_x$ and $S_y$ acts as a discrete PH transformation. For single-particle Hamiltonian $H$ this leads to
\begin{equation}
\sigma_y H^T \sigma_y = - H.
\end{equation}
The Hamiltonians with this constraints form the symmetry class C. Now by imposing TR symmetry, which combined with PH symmetry gives,
\begin{equation}
\sigma_y H^T \sigma_y = - H,  \;\;\;\;\;\;\;\;\; H^*=H.
\end{equation}
The BdG Hamiltonians with this condition form symmetry class CI.

\section{Model Hamiltonians}
In this theoretical study we present all BdG Hamiltonians in the momentum space. Also all the Hamiltonians preserve hermiticity property ($\mathcal{H}(k)=\mathcal{H}^{\dagger}(k)$).\\
\textbf{First Hamiltonian:} We consider the first BdG Hamiltonian, which belongs to the study of topological state of matter, i.e. the Kitaev chain in momentum space \cite{kitaev2001}. Its Bloch Hamiltonian has the matrix form 
\begin{equation}
\mathcal{H}^{(1)}(k)= \left( \begin{matrix}
	\epsilon_k - \mu && \Delta_k\\
	\Delta^*_k && -\epsilon_k + \mu \\
	\end{matrix}\right), \label{kitaev}
\end{equation}
where $\epsilon_k=-2t\cos k$ with $t$ as hopping matrix element, $\Delta_k=2i\Delta\sin(k)$ with $\Delta$ as superconducting gap, and $\mu$ is chemical potential. We study the symmetries of this Hamiltonian and observe which symmetry class the Hamiltonian belongs to in Table \ref{periodictable}.\\ 
\textbf{Second Hamiltonian}: Here we consider BdG Hamiltonian with a variant term $\alpha k$ added in the $\sigma_x$ component of the Hamiltonian
\begin{equation}
\mathcal{H}^{(2)}(k)=\left( \begin{matrix}
\epsilon_k-\mu - \alpha k &&  \Delta_k \\
\Delta^*_k && -\epsilon_k+\mu + \alpha k
\end{matrix}\right).
\end{equation}
Here variant term is added to the component of $\sigma_x$ of the Hamiltonian.\\
\textbf{Third Hamiltonian}: Here we consider BdG Hamiltonian with the variant term added in the component of $\sigma_y$ of the Hamiltonian
 \begin{equation}
 \mathcal{H}^{(3)}(k) =\left( \begin{matrix}
\epsilon_k-\mu  &&  \Delta_k + i\alpha k\\
\Delta^*_k-i\alpha k && -\epsilon_k+\mu
\end{matrix}\right).
 \end{equation} 
Here variant term is added to the potential component of the model Hamiltonian.\\
\noindent \textbf{Fourth Hamiltonian}: Here we consider BdG Hamiltonian where the variant term is added to both the components, i.e. in the $\sigma_x$ component and also in the $\sigma_y$ component of the Hamiltonian. Hamiltonian takes the form
\begin{equation}
 \mathcal{H}^{(4)}(k)=\left( \begin{matrix}
\epsilon_k-\mu-\alpha k  &&  \Delta_k + i\alpha k\\
\Delta^*_k-i\alpha k && -\epsilon_k+\mu+\alpha k
\end{matrix}\right). 
\end{equation}
All the three Hamiltonians with the variant term preserve the Hermiticity of the system. Here $\alpha$ is a real number (either positive or negative) and $k$ is the momentum.  Since the variant Hamiltonian is linear in momentum ($\alpha k$ - variant term $\alpha$ is linear in momentum $k$), it will be suitable for quantum simulation process. 
The physics of quantum simulation is state of the art to simulate different kinds of quantum many-body systems \cite{sarkar2nature,Katsura,Fidkowski,Kitaev-int}. Thus the variant term above may help in effective quantum simulation of the system. At first we consider the variant Hamiltonians in spinless background.\\
\section{Results of symmetries for BdG Hamiltonians in spinless background}
Here we present results of symmetry study for all BdG Hamiltonians in spinless background.
At first we discuss the basic symmetry aspects for the Hamiltonian $ \mathcal{H}^{(1)}(k)$. We can show
\begin{equation*}
          \mathcal{T}\mathcal{H}^{(1)}(k)\mathcal{T}^{-1} = \mathcal{H}^{(1)}(k) ,
          \;\;\;\;   \mathcal    {C}\mathcal{H}^{(1)}(k)\mathcal{C}^{-1}   =
          -\mathcal{H}^{(1)}(k),
 	\end{equation*} 
 	\begin{equation} \mathcal{S}\mathcal{H}^{(1)}(k)\mathcal{S}^{-1} = -\mathcal{H}^{(1)}(k).
 \end{equation}
Thus the Hamiltonian obeys time-reversal and particle-hole symmetries. 
One can write the BdG Hamiltonian  (eq.\ref{kitaev}) in the spin basis as
\begin{equation}
\mathcal{H}^{(1)}(k)=b(k).\tau,
\end{equation}
where $b(k)$ is an effective Zeeman field and $\tau$ is Pauli spin matrices in particle-hole space. From eq.10 we have the components $b_x(k)=0$, $b_y(k)=2i\Delta\sin(k)$ and $b_z(k)=-2t\cos(k)-\mu$. Only two of the three Pauli matrices are involved here, thus one can have an anti-commutation relation between $\sigma_x$ and $\mathcal{H}^{(1)}(k)$, $\left\lbrace \sigma_x,\mathcal{H}^{(1)}(k) \right\rbrace =0$. This defines the chiral symmetry with the operator $\sigma_x$\cite{von2017topological,kane2013topological}.
\begin{equation}
\left\lbrace \sigma_x,\mathcal{H}^{(1)}(k)  \right\rbrace = \sigma_x\mathcal{H}^{(1)}(k)+\mathcal{H}^{(1)}(k)\sigma_x=0
\end{equation}
Now we check for the condition
\begin{equation}
\sigma_x\mathcal{H}^{(1)}(k)=-\mathcal{H}^{(1)}(k)\sigma_x
\end{equation}
\begin{equation}
\begin{aligned}
\sigma_x\mathcal{H}^{(1)}(k)&= \left( \begin{matrix}
0 && 1\\
1 && 0\\
\end{matrix}\right) \left( \begin{matrix}
\epsilon_{k,\sigma}-\mu && \Delta_k\\
\Delta_k^* && -\epsilon_{k,\sigma}+\mu
\end{matrix}\right) \\
&= \left( \begin{matrix}
\Delta_k^* && -\epsilon_{k,\sigma}+\mu\\
\epsilon_{k,\sigma}-\mu && \Delta_k 
\end{matrix}\right) .
\end{aligned}
\end{equation}
Similarly,
\begin{equation}
\begin{aligned}
\mathcal{H}^{(1)}(k)\sigma_x&= \left( \begin{matrix}
\epsilon_{k,\sigma}-\mu && \Delta_k\\
\Delta_k^* && -\epsilon_{k,\sigma}+\mu
\end{matrix}\right) \left( \begin{matrix}
0 && 1\\
1 && 0\\
\end{matrix}\right) \\
&= \left( \begin{matrix}
\Delta_k && \epsilon_{k,\sigma}-\mu\\
-\epsilon_{k,\sigma}+\mu && \Delta_k^*
\end{matrix}\right) .
\end{aligned}
\end{equation}
Thus we have
\begin{equation}
\sigma_x\mathcal{H}^{(1)}(k)=-\mathcal{H}^{(1)}(k)\sigma_x.
\end{equation}
A similar calculation for $\sigma_y$ and  $\sigma_z$ gives
\begin{equation*}
\sigma_y\mathcal{H}^{(1)}(k)\neq-\mathcal{H}^{(1)}(k)\sigma_y ,
\end{equation*}
\begin{equation}
\sigma_z\mathcal{H}^{(1)}(k)\neq-\mathcal{H}^{(1)}(k)\sigma_z.
\end{equation}
We observe that  the Hamiltonian also obeys the condition for chiral symmetry which anti-commutes with the Hamiltonian. Thus $\mathcal{H}^{(1)}(k)$ satisfies the condition for all the three symmetries and falls under the symmetry class BDI of Table 1. The first row of Table 2 shows the symmetry table for the Hamiltonian $\mathcal{H}^{(1)}(k)$. $\mathcal{Z}$ is topological invariant number which takes integer values. Each value of $\mathcal{Z}$ indicates a set of $\mathcal{H}^{(1)}(k)$ which can be interpolated continuously without breaking the symmetries and without closing the energy gap. One can tune the parameters of $\mathcal{H}^{(1)}(k)$ to get gapless state. This closing the gap involves increase in $\mathcal{Z}$ by one unit and it defines topologically distinct set of $\mathcal{H}^{(1)}(k)$. \\
\begin{table}
\begin{center}
\begin{tabular}{|c|c|c|c|}
\hline
 & Symmetry & dimension (d) \\

& AZ \;\;\; $\mathcal{T}^2$ \;\;\; $\mathcal{C}^2$ \;\;\; $\mathcal{S}^2$ & 1 \\
\hline
$\mathcal{H}^{(1)}(k)$ & BDI \;\;\; 1 \;\;\;\;\; 1 \;\;\;\;\; 1 & $\mathbb{Z}$ \\
\hline
$\mathcal{H}^{(2)}(k)$ & AIII \;\;\; 0 \;\;\;\;\; 0 \;\;\;\;\; 1 & $\mathbb{Z}$ \\
\hline
$\mathcal{H}^{(3)}(k)$ & BDI \;\;\; 1 \;\;\;\;\; 1 \;\;\;\;\; 1 & $\mathbb{Z}$ \\
\hline
$\mathcal{H}^{(4)}(k)$ & AIII \;\;\; 0 \;\;\;\;\; 0 \;\;\;\;\; 1 & $\mathbb{Z}$ \\
\hline
\end{tabular} 
\end{center}
\caption{Symmetry table for the BdG model Hamiltonians in spinless background.}
\end{table}
Now we present symmetry studies for the Hamiltonian $\mathcal{H}^{(2)}(k)$. We can show
 \begin{equation*}
  \mathcal{T}\mathcal{H}^{(2)}(k)\mathcal{T}^{-1}=\mathcal{H}^{(2)}(k),
  \;\;\;    \mathcal{C}\mathcal{H}^{(2)}(k)\mathcal{C}^{-1}     =    -
  \mathcal{H}^{(2)}(k),
 \end{equation*}
 \begin{equation}
 \mathcal{S}\mathcal{H}^{(2)}(k)\mathcal{S}^{-1}=                    -
 \mathcal{H}^{(2)}(k).
 \end{equation}
Here we observe that the Hamiltonian $\mathcal{H}^{(2)}(k)$ satisfies the condition for only chiral symmetry and it does not hold TR and PH symmetries. Thus it falls under the symmetry class AIII of Table 1. \\
Now we present symmetry studies for the Hamiltonian $\mathcal{H}^{(3)}(k)$. We can show
 \begin{equation*}
  \mathcal{T}\mathcal{H}^{(3)}(k)\mathcal{T}^{-1}\neq\mathcal{H}^{(3)}(k),
  \;\;\;   \mathcal{C}\mathcal{H}^{(3)}(k)\mathcal{C}^{-1}    \neq   -
  \mathcal{H}^{(3)}(k),
 \end{equation*}
 \begin{equation}
 \mathcal{S}\mathcal{H}^{(3)}(k)\mathcal{S}^{-1}=                    -
 \mathcal{H}^{(3)}(k).
 \end{equation}
Here we  observe that the Hamiltonian $\mathcal{H}^{(3)}(k)$ satisfies the condition for all the three symmetries. Thus the Hamiltonian falls under the symmetry class BDI of Table 1.\\
Now we present symmetry studies for the Hamiltonian $\mathcal{H}^{(4)}(k)$. We can show
\begin{equation*}
  \mathcal{T}\mathcal{H}^{(4)}(k)\mathcal{T}^{-1}\neq\mathcal{H}^{(4)}(k),
  \;\;\;   \mathcal{C}\mathcal{H}^{(4)}(k)\mathcal{C}^{-1}    \neq   -
  \mathcal{H}^{(4)}(k),
 \end{equation*}
 \begin{equation}
 \mathcal{S}\mathcal{H}^{(4)}(k)\mathcal{S}^{-1}=                    -
 \mathcal{H}^{(4)}(k).
 \end{equation}
We observe that the Hamiltonian $\mathcal{H}^{(4)}(k)$	satisfies the symmetry condition only in the case of chiral symmetry. Thus it falls under the symmetry class AIII. Table 2 summarises the results of symmetry studies for model Hamiltonians for spinless background. We have $\mathcal{H}^{(1)}(k)$ and $\mathcal{H}^{(3)}(k)$ in BDI symmetry class and $\mathcal{H}^{(2)}(k)$ and $\mathcal{H}^{(4)}(k)$ in AIII symmetry class, both with $\mathcal{Z}$ topological invariant number.	Thus it is clear from this study that the variant of $\sigma_x$ and $\sigma_y$ components of the BdG Hamiltonian leads to different symmetry classes.\\



\section{Results of symmetries for BdG Hamiltonians in spinful background}
\begin{figure}[h]
						{\includegraphics[scale=0.43]{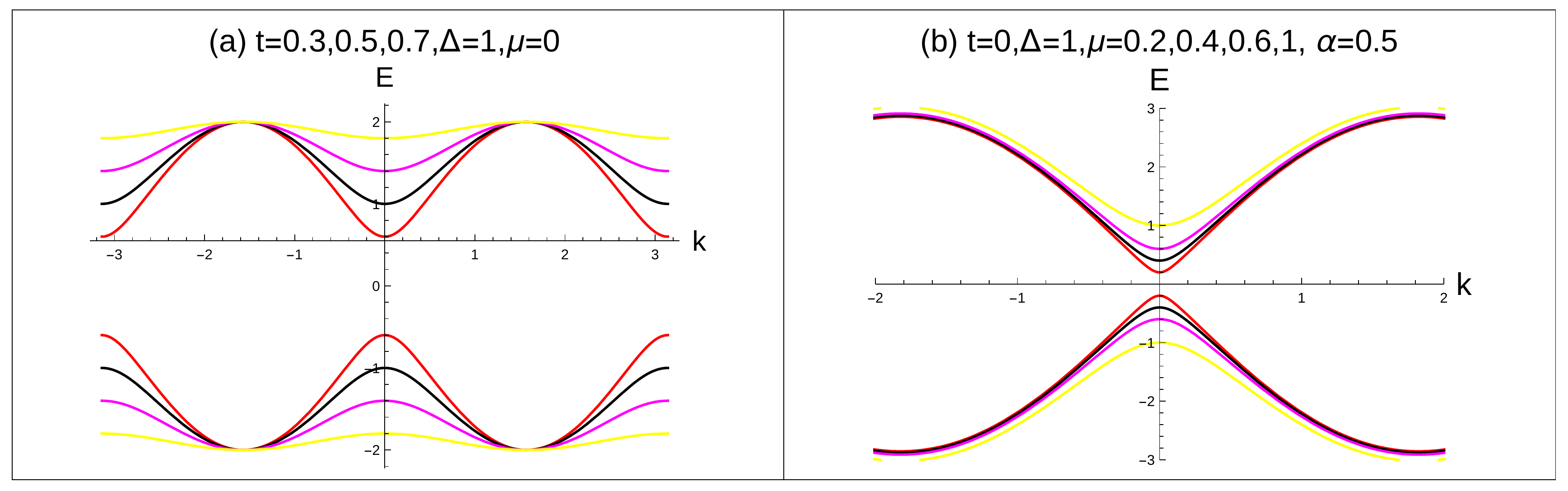}}
						\caption{(a) Energy dispersion for $\mathcal{H}^{(1)}(k,\sigma)$ for different values of $t$, 
					    (b) Energy dispersion for $\mathcal{H}^{(3)}(k,\sigma)$ for different values of $\mu$.}\label{dis1} 
\end{figure}

Here we study BdG Hamiltonian in spinful background. It is well known to us that the symmetry operators for TR symmetry is different for the spinless and spinful background. We also try to find there is any emergence of extra symmetry class. We consider the Hamiltonians $\mathcal{H}^{(1)}(k)$, $\mathcal{H}^{(2)}(k)$, $\mathcal{H}^{(3)}(k)$ and $\mathcal{H}^{(4)}(k)$ as spinful systems with symmetry operators $\mathcal{T}=i\sigma_y {K}$ for TR, $\mathcal{C}=\sigma_x {K}$ for PH and $\mathcal{S}=\sigma_x$ for chiral symmetries. One may also consider the BdG Hamiltonians in the form of the spinful fermionic system as,\\
\begin{equation}
\mathcal{H}^{(1)}(k,\sigma)= \left( \begin{matrix}
	\epsilon_{k,\sigma} - \mu && \Delta_k\\
	\Delta^*_k && -\epsilon_{k,\sigma} + \mu \\
	\end{matrix}\right), 
\mathcal{H}^{(2)}(k,\sigma)=\left( \begin{matrix}
	\epsilon_{k,\sigma}-\mu - \alpha k &&  \Delta_k \\
	\Delta^*_k && -\epsilon_{k,\sigma}+\mu + \alpha k
	\end{matrix}\right),
\end{equation}
 \begin{equation}
\mathcal{H}^{(3)}(k,\sigma) =\left( \begin{matrix}
\epsilon_{k,\sigma}-\mu  &&  \Delta_k + i\alpha k\\
\Delta^*_k-i\alpha k && -\epsilon_{k,\sigma}+\mu
\end{matrix}\right),
\mathcal{H}^{(4)}(k,\sigma)=\left( \begin{matrix}
\epsilon_{k,\sigma}-\mu-\alpha k  &&  \Delta_k + i\alpha k\\
\Delta^*_k-i\alpha k && -\epsilon_{k,\sigma}+\mu+\alpha k
\end{matrix}\right),
 \end{equation} 
 where $\epsilon_{k,\sigma}=-2t\cos k$, and $t$ is spin-independent hopping term.
Now we consider  $\mathcal{H}^{(1)}(k,\sigma)$ and study its symmetry properties in spinful background. We check for invariance of the model Hamiltonians under the TR, PH and chiral symmetry operations.\\
\begin{equation}
\begin{aligned}
	 \mathcal{T}\mathcal{H}^{(1)}(k,\sigma)\mathcal{T}^{-1} &= (i\sigma_y {K}) \mathcal{H}^{(1)}(k,\sigma) (i\sigma_y {K})^{-1} =  (i\sigma_y) {K}\mathcal{H}^{(1)}(k,\sigma) {K}(-i\sigma_y)\\
	 &=  \left( \begin{matrix}
	 0 && -1\\
	 1 && 0\\
	 \end{matrix}\right) \left(  \begin{matrix}
	 \epsilon_{k,\sigma}-\mu && 2i\Delta\sin k\\
	 -2i\Delta\sin k && -\epsilon_{k,\sigma}+\mu\\
	 \end{matrix}\right) \left(  \begin{matrix}
	 0 && 1\\
	 -1 && 0\\
	 \end{matrix}\right) \\
	 &= \left( \begin{matrix}
	 -\epsilon_{k,\sigma}+\mu && 2i\Delta\sin k\\
	 -2i\Delta\sin k && \epsilon_{k,\sigma}-\mu\\
	 \end{matrix}\right) \\
	 &\neq \mathcal{H}^{(1)}(k,\sigma).
	\end{aligned}
\end{equation}
\begin{equation}
\begin{aligned}
	\mathcal{C}\mathcal{H}^{(1)}(k,\sigma)\mathcal{C}^{-1} &= (\sigma_x {K}) \mathcal{H}^{(1)}(k,\sigma) (\sigma_x {K})^{-1}
		=  \sigma_x {K}\mathcal{H}^{(1)}(k,\sigma) {K}\sigma_x\\
		&=   \left( \begin{matrix}
		0 && 1\\
		1 && 0\\
		\end{matrix}\right)  \left( \begin{matrix}
		\epsilon_{k,\sigma}-\mu && 2i\Delta\sin k\\
		-2i\Delta\sin k && -\epsilon_{k,\sigma}+\mu\\
		\end{matrix}\right)  \left( \begin{matrix}
		0 && 1\\
		1 && 0\\
		\end{matrix}\right)\\
		&= \left( \begin{matrix}
	-\epsilon_{k,\sigma}+\mu && -2i\Delta\sin k\\
		2i\Delta\sin k && \epsilon_{k,\sigma}-\mu\\
		\end{matrix}\right)\\
		&= - \mathcal{H}^{(1)}(k,\sigma).
		\end{aligned}
	\end{equation}
\begin{equation}
\begin{aligned}
	\mathcal{S}\mathcal{H}^{(1)}(k,\sigma)\mathcal{S}^{-1} &= (\sigma_x) \mathcal{H}^{(1)}(k,\sigma) (\sigma_x)^{-1}\\
	 &= \left( \begin{matrix}
	 0 && 1\\
	 1 && 0\\
	 \end{matrix}\right) \left(  \begin{matrix}
	 \epsilon_{k,\sigma}-\mu && 2i\Delta\sin k\\
	 -2i\Delta\sin k && -\epsilon_{k,\sigma}+\mu\\
	 \end{matrix}\right)  \left( \begin{matrix}
	 0 && 1\\
	 1 && 0\\
	 \end{matrix}\right) \\
	 &= \left( \begin{matrix}
	 -\epsilon_{k,\sigma}+\mu && -2i\Delta\sin k\\
	 2i\Delta\sin k && \epsilon_{k,\sigma}-\mu\\
	 \end{matrix}\right) \\
	 &= - \mathcal{H}^{(1)}(k,\sigma).
\end{aligned}
\end{equation}
Here we observe that the Hamiltonian $\mathcal{H}^{(1)}(k,\sigma)$	satisfies the condition for both PH and chiral symmetry but not TR symmetry.
\begin{equation*}
  \mathcal{T}\mathcal{H}^{(1)}(k,\sigma)\mathcal{T}^{-1}\neq\mathcal{H}^{(1)}(k,\sigma),
  \;\;\;   \mathcal{C}\mathcal{H}^{(1)}(k,\sigma)\mathcal{C}^{-1}   =  -
  \mathcal{H}^{(1)}(k,\sigma),
 \end{equation*}
 \begin{equation}
 \mathcal{S}\mathcal{H}^{(1)}(k,\sigma)\mathcal{S}^{-1}=                    -
 \mathcal{H}^{(1)}(k,\sigma).
 \end{equation}
This is an interesting result since this set of symmetry does not belong to any symmetry class of Table 1. The Hamiltonian $\mathcal{H}^{(1)}(k,\sigma)$ has gapless state for the condition $\mu=-2t$ at $k=0$ and also for the condition $\mu=2t$ at $k=\pm\pi$. From the energy dispersion curve (fig.\ref{dis1}a) we can observe that the energy spectrum is symmetric about zero. This is clearly one of the implications of PH and chiral symmetries.
In Table 1 one can observe presence of chiral symmetry only in the presence or absence of both TR and PH symmetries. There is no symmetry class with either TR or PH along with chiral symmetry. But this symmetry class has PH symmetry along with chiral symmetry. 
\begin{table}
\begin{center}
\begin{tabular}{|c|c|c|c|}
\hline
 & Symmetry & dimension (d) \\

& AZ \;\;\;\;\; $\mathcal{T}^2$ \;\;\; $\mathcal{C}^2$ \;\;\; $\mathcal{S}^2$ & 1 \\
\hline
$\mathcal{H}^{(1)}(k)$ & X   \;\;\;\;\;\;\; 0 \;\;\;\;\; 1 \;\;\;\;\; 1 & $\mathcal{X}$ \\
\hline
$\mathcal{H}^{(2)}(k)$ & AIII \;\;\; 0 \;\;\;\;\; 0 \;\;\;\;\; 1 & $\mathbb{Z}$ \\
\hline
$\mathcal{H}^{(3)}(k)$ & X   \;\;\;\;\;\;\; 0 \;\;\;\;\; 1 \;\;\;\;\; 1 & $\mathcal{X}$ \\
\hline
$\mathcal{H}^{(4)}(k)$ & AIII \;\;\; 0 \;\;\;\;\; 0 \;\;\;\;\; 1 & $\mathbb{Z}$ \\
\hline
\end{tabular} 
\end{center}
\caption{Symmetry table for model BdG Hamiltonians in spinful background.}
\end{table}
Now we consider  $\mathcal{H}^{(2)}(k,\sigma)$ and study its symmetry properties.
\begin{equation}
\begin{aligned}
	 \mathcal{T}\mathcal{H}^{(2)}(k,\sigma)\mathcal{T}^{-1} &= (i\sigma_y {K}) \mathcal{H}^{(2)}(k,\sigma) (i\sigma_y {K})^{-1} =  (i\sigma_y) {K}\mathcal{H}^{(2)}(k,\sigma) {K}(-i\sigma_y)\\
	 &=  \left( \begin{matrix}
	 0 && -1\\
	 1 && 0\\
	 \end{matrix}\right) \left(  \begin{matrix}
	 \epsilon_{k,\sigma}-\mu+\alpha k && 2i\Delta\sin k\\
	 -2i\Delta\sin k && -\epsilon_{k,\sigma}+\mu-\alpha k\\
	 \end{matrix}\right) \left(  \begin{matrix}
	 0 && 1\\
	 -1 && 0\\
	 \end{matrix}\right) \\
	 &= \left( \begin{matrix}
	 -\epsilon_{k,\sigma}+\mu-\alpha k && 2i\Delta\sin k\\
	 -2i\Delta\sin k && \epsilon_{k,\sigma}-\mu+\alpha k\\
	 \end{matrix}\right) \\
	 &\neq \mathcal{H}^{(2)}(k,\sigma).
	\end{aligned}
\end{equation}
\begin{equation}
\begin{aligned}
	\mathcal{C}\mathcal{H}^{(2)}(k,\sigma)\mathcal{C}^{-1} &= (\sigma_x {K}) \mathcal{H}^{(2)}(k,\sigma) (\sigma_x {K})^{-1}
		=  \sigma_x {K}\mathcal{H}^{(2)}(k,\sigma) {K}\sigma_x\\
		&=   \left( \begin{matrix}
		0 && 1\\
		1 && 0\\
		\end{matrix}\right)  \left( \begin{matrix}
		\epsilon_{k,\sigma}-\mu+\alpha k && 2i\Delta\sin k\\
		-2i\Delta\sin k && -\epsilon_{k,\sigma}+\mu-\alpha k\\
		\end{matrix}\right)  \left( \begin{matrix}
		0 && 1\\
		1 && 0\\
		\end{matrix}\right)\\
		&= \left( \begin{matrix}
		-\epsilon_{k,\sigma}+\mu-\alpha k && -2i\Delta\sin k\\
		2i\Delta\sin k && \epsilon_{k,\sigma}-\mu+\alpha k\\
		\end{matrix}\right)\\
		&\neq - \mathcal{H}^{(2)}(k,\sigma).
		\end{aligned}
	\end{equation}
\begin{equation}
\begin{aligned}
	\mathcal{S}\mathcal{H}^{(2)}(k,\sigma)\mathcal{S}^{-1} &= (\sigma_x) \mathcal{H}^{(2)}(k,\sigma) (\sigma_x)^{-1}\\
	 &= \left( \begin{matrix}
	 0 && 1\\
	 1 && 0\\
	 \end{matrix}\right) \left(  \begin{matrix}
	 \epsilon_{k,\sigma}-\mu-\alpha k && 2i\Delta\sin k\\
	 -2i\Delta\sin k && -\epsilon_{k,\sigma}+\mu+\alpha k\\
	 \end{matrix}\right)  \left( \begin{matrix}
	 0 && 1\\
	 1 && 0\\
	 \end{matrix}\right) \\
	 &= \left( \begin{matrix}
	 -\epsilon_{k,\sigma}+\mu+\alpha k && -2i\Delta\sin k\\
	 2i\Delta\sin k && \epsilon_{k,\sigma}-\mu-\alpha k\\
	 \end{matrix}\right) \\
	 &= - \mathcal{H}^{(2)}(k,\sigma).
\end{aligned}
\end{equation}
Here we observe that the Hamiltonian $\mathcal{H}^{(2)}(k,\sigma)$	satisfies the symmetry condition only in the case of chiral symmetry. 
\begin{equation*}
  \mathcal{T}\mathcal{H}^{(2)}(k,\sigma)\mathcal{T}^{-1}\neq\mathcal{H}^{(2)}(k,\sigma),
  \;\;\;   \mathcal{C}\mathcal{H}^{(2)}(k,\sigma)\mathcal{C}^{-1}    \neq   -
  \mathcal{H}^{(2)}(k,\sigma),
\end{equation*}
\begin{equation}
 \mathcal{S}\mathcal{H}^{(2)}(k,\sigma)\mathcal{S}^{-1}=                    -
 \mathcal{H}^{(2)}(k,\sigma).
\end{equation}
Thus it falls under the symmetry class AIII.


Now we consider the Hamiltonian $\mathcal{H}^{(3)}(k,\sigma)$ and study its symmetry properties. We observe the appearance of an extra symmetry class as in the case of $\mathcal{H}^{(1)}(k,\sigma)$ with same symmetry configuration.
\begin{equation}
\begin{aligned}
	\mathcal{T}\mathcal{H}^{(3)}(k,\sigma)\mathcal{T}^{-1}&= (i\sigma_y {K}) \mathcal{H}^{(3)}(k,\sigma) (i\sigma_y {K})^{-1}.
     = (i\sigma_y){K} \mathcal{H}^{(3)}(k,\sigma) {K} (-i\sigma_y).\\
	 &=  \left(\begin{matrix}
	 0 && -1\\
	 1 && 0
	 \end{matrix}\right) \left(\begin{matrix}
	 \epsilon_{k,\sigma}-\mu && 2i\Delta\sin k+i\alpha k\\
	 -2i\Delta\sin k-i\alpha k && -\epsilon_{k,\sigma}+\mu
	 \end{matrix} \right) \left(\begin{matrix}
	 0 && 1\\
	 -1 && 0
	 \end{matrix}\right) \\ 
	 &= \left(\begin{matrix}
	 -\epsilon_{k,\sigma}+\mu && 2i\Delta\sin k+i\alpha k\\
	 -2i\Delta\sin k-i\alpha k && \epsilon_{k,\sigma}-\mu
	 \end{matrix}\right)\\
	 & \neq \mathcal{H}^{(3)}(k,\sigma).
\end{aligned}
\end{equation}	 
\begin{equation}
\begin{aligned}
	 \mathcal{C}\mathcal{H}^{(3)}(k,\sigma)\mathcal{C}^{-1} &= (\sigma_x{K})\mathcal{H}^{(3)}(k,\sigma)(\sigma_x {K})^{-1}
	 	= \sigma_x{K}\mathcal{H}^{(3)}(k,\sigma){K}\sigma_x\\
	 	&= \left(\begin{matrix}
	 	0 && 1\\
	 	1 &&  0 \\
	 	\end{matrix}\right) \left(\begin{matrix}
	 	\epsilon_{k,\sigma}-\mu  && 2i\Delta\sin k+i\alpha k\\
	 	-2i\Delta\sin k-i\alpha k && -\epsilon_{k,\sigma}+\mu\\
	 	\end{matrix}\right) \left(\begin{matrix}
	 	0 && 1\\
	 	1 &&  0 \\
	 	\end{matrix}\right)\\
	 	&=\left(\begin{matrix}
	 	-\epsilon_{k,\sigma}+\mu  && -2i\Delta\sin k-i\alpha k\\
	 	2i\Delta\sin k+i\alpha k && \epsilon_{k,\sigma}-\mu\\
	 	\end{matrix}\right)\\
	 	& = -\mathcal{H}^{(3)}(k,\sigma).
	 \end{aligned}
	 \end{equation}
\begin{equation}
\begin{aligned}
	\mathcal{S}\mathcal{H}^{(3)}(k,\sigma)\mathcal{S}^{-1} &= \sigma_x \mathcal{H}^{(3)}(k,\sigma)\sigma_x^{-1} \\
	 &= \left(\begin{matrix}
	 0 && 1\\
	 1 &&  0 \\
	 \end{matrix}\right) \left(\begin{matrix}
	 \epsilon_{k,\sigma}-\mu  && 2i\Delta\sin k+i\alpha k\\
	 -2i\Delta\sin k-i\alpha k && -\epsilon_{k,\sigma}+\mu\\
	 \end{matrix}\right) \left(\begin{matrix}
	 0 && 1\\
	 1 &&  0 \\
	 \end{matrix}\right)\\
	 &= \left(\begin{matrix}
	 -\epsilon_{k,\sigma}+\mu  && -2i\Delta\sin k-i\alpha k\\
	 2i\Delta\sin k+i\alpha k && \epsilon_{k,\sigma}-\mu\\
	 \end{matrix}\right)\\
	 &= -\mathcal{H}^{(3)}(k,\sigma).
\end{aligned}
\end{equation}
Here we observe that the Hamiltonian $\mathcal{H}^{(3)}(k,\sigma)$	also satisfies the condition for both PH and chiral symmetry but not for TR symmetry.
\begin{equation*}
  \mathcal{T}\mathcal{H}^{(3)}(k,\sigma)\mathcal{T}^{-1}\neq\mathcal{H}^{(3)}(k,\sigma),
  \;\;\;   \mathcal{C}\mathcal{H}^{(3)}(k,\sigma)\mathcal{C}^{-1}  =  -
  \mathcal{H}^{(3)}(k,\sigma),
 \end{equation*}
 \begin{equation}
 \mathcal{S}\mathcal{H}^{(3)}(k,\sigma)\mathcal{S}^{-1}=                    -
 \mathcal{H}^{(3)}(k,\sigma).
 \end{equation}
This symmetry class is same as in  $\mathcal{H}^{(1)}(k,\sigma)$ and does not belong to any symmetry class of Table 1. The Hamiltonian $\mathcal{H}^{(3)}(k,\sigma)$ has gapless state only for the condition $\mu=-2t$ at $k=0$ and there is no gapless state for $k=\pm\pi$. Similar to the case of $\mathcal{H}^{(1)}(k,\sigma)$ here also from the energy dispersion curve (fig.\ref{dis1}b) we can observe that the energy spectrum is symmetric about zero. This is clearly one of the implications of PH and chiral symmetries.
In Table 1 one can observe presence of chiral symmetry only in presence or absence of both TR and PH symmetries. There is no symmetry class with either TR or PH along with chiral symmetry. But this symmetry class has PH symmetry along with chiral symmetry. Therefore this study predict that even though the Hamiltonians are the same, spinless and spinful backgrounds sometimes do not give the same result. \\	

Now we consider the Hamiltonian $\mathcal{H}^{(4)}(k,\sigma)$ and study its symmetry properties.
\begin{equation}
\begin{aligned}
	 \mathcal{T}\mathcal{H}^{(4)}(k,\sigma)\mathcal{T}^{-1} &= (i\sigma_y {K}) \mathcal{H}^{(4)}(k,\sigma) (i\sigma_y {K})^{-1}
	 =  (i\sigma_y){K} \mathcal{H}^{(4)}(k,\sigma) {K}(-i\sigma_y)\\
	 &=  \left( \begin{matrix}
	 0 && -1\\
	 1 && 0\\
	 \end{matrix}\right) \left(  \begin{matrix}
	 \epsilon_{k,\sigma}-\mu+\alpha k && 2i\Delta\sin k+i\alpha k\\
	 -2i\Delta\sin k-i\alpha k && -\epsilon_{k,\sigma}+\mu-\alpha k \\
	 \end{matrix} \right) \left( \begin{matrix}
	 0 && 1\\
	 -1 && 0\\
	 \end{matrix}\right)\\
	 &= \left( \begin{matrix}
	 -\epsilon_{k,\sigma}+\mu-\alpha k && 2i\Delta\sin k+i\alpha k\\
	 -2i\Delta\sin k-i\alpha k && \epsilon_{k,\sigma}-\mu+\alpha k \\
	 \end{matrix}\right) \\
	 &\neq \mathcal{H}^{(4)}(k,\sigma).
\end{aligned}
\end{equation}
\begin{equation}
\begin{aligned}
     \mathcal{C}\mathcal{H}^{(4)}(k,\sigma)\mathcal{C}^{-1} &= (\sigma_x {K}) \mathcal{H}^{(4)}(k,\sigma) (\sigma_x {K})^{-1}
     	= \sigma_x {K}\mathcal{H}^{(4)}(k,\sigma) {K}\sigma_x\\
     	&=  \left( \begin{matrix}
     	0 && 1\\
     	1 && 0\\
     	\end{matrix}\right)  \left( \begin{matrix}
     	\epsilon_{k,\sigma}-\mu+\alpha k && 2i\Delta\sin k+i\alpha k\\
     	-2i\Delta\sin k-i\alpha k && -\epsilon_{k,\sigma}+\mu-\alpha k \\
     	\end{matrix}\right) \left(  \begin{matrix}
     	0 && 1\\
     	1 && 0\\
     	\end{matrix}\right)\\
     	&=\left( \begin{matrix}
     	-\epsilon_{k,\sigma}+\mu-\alpha k && -2i\Delta\sin k-i\alpha k\\
     	2i\Delta\sin k+i\alpha k && \epsilon_{k,\sigma}-\mu+\alpha k \\
     	\end{matrix}\right)\\
     	&\neq - \mathcal{H}^{(4)}(k,\sigma).
     \end{aligned}
     \end{equation}
	\begin{equation}
	\begin{aligned}
	\mathcal{S}\mathcal{H}^{(4)}(k,\sigma)\mathcal{S}^{-1} &= (\sigma_x) \mathcal{H}^{(4)}(k,\sigma) (\sigma_x)^{-1}\\
	&= \left( \begin{matrix}
	0 && 1\\
	1 && 0\\
	\end{matrix}\right) \left(  \begin{matrix}
	\epsilon_{k,\sigma}-\mu-\alpha k && 2i\Delta\sin k+i\alpha k\\
	-2i\Delta\sin k-i\alpha k && -\epsilon_{k,\sigma}+\mu+\alpha k \\
	\end{matrix}\right)  \left( \begin{matrix}
	0 && 1\\
	1 && 0\\
	\end{matrix}\right) \\
	&= - \mathcal{H}^{(4)}(k,\sigma).
	\end{aligned}
	\end{equation}
Here we observe that the Hamiltonian $\mathcal{H}^{(4)}(k,\sigma)$	satisfies the symmetry condition only in the case of chiral symmetry.
\begin{equation*}
  \mathcal{T}\mathcal{H}^{(4)}(k,\sigma)\mathcal{T}^{-1}\neq\mathcal{H}^{(4)}(k,\sigma),
  \;\;\;   \mathcal{C}\mathcal{H}^{(4)}(k,\sigma)\mathcal{C}^{-1}    \neq   -
  \mathcal{H}^{(4)}(k,\sigma),
 \end{equation*}
 \begin{equation}
 \mathcal{S}\mathcal{H}^{(4)}(k,\sigma)\mathcal{S}^{-1}=                    -
 \mathcal{H}^{(4)}(k,\sigma).
 \end{equation} 
Thus it falls under the symmetry class AIII. Table III shows the results of symmetry studies of model BdG Hamiltonians in spinful background. A new symmetry class appears for $\mathcal{H}^{(1)}(k,\sigma)$ and $\mathcal{H}^{(3)}(k,\sigma)$, whereas $\mathcal{H}^{(2)}(k,\sigma)$ and $\mathcal{H}^{(4)}(k,\sigma)$ fall under AIII symmetry class as in spinless background.	\\
We know according to Kramers's theorem that a spinful system which is TR invariant has atleast two-fold degenerate ground state, which are called Kramers's degeneracy.  This theorem is a direct consequence of the property $\mathcal{T}^2=-1$. But there is no possibility for the appearence of Kramers's degeneracy for our model Hamiltonians in spinless background since time-reversal operator has property $\mathcal{T}^2=1$. However, in the case of spinful background one could expect the Kramers's degeneracy since $\mathcal{T}^2=-1$. But in our case none of the model Hamiltonians preserves time-reversal symmetry in the spinful background. Thus we observe no evidence of the Kramers's degeneracy. 
	 				


\section{Conclusion}
We have done extensive study of the symmetries for a set of BdG model Hamiltonians. We observe that for spinful case the Hamiltonians $\mathcal{H}^{(1)}(k,\sigma)$ and $\mathcal{H}^{(3)}(k,\sigma)$ show a new symmetry class, which is beyond the ten-fold symmetry classes. Our strong belief is that quantum simulation scientists will be motivated by this study to find these new types of BdG model Hamiltonians. We also found that the model Hamiltonians are not TR invariant under the spinful background and also not show Kramers's degeneracy.
\section*{Acknowledgment}
S.S. would like to acknowledge DST (EMR/2017/000898) for support. R.K.R. and S.S. would like to acknowledge Raman Research Institute library for books and journals and also the academic activities of International Centre for Theoretical Sciences. R.K.R. and S.S. would also like to acknowledge Mr. N.A. Prakash for critical reading of this manuscript.\\

\bibliography{firstpaper}
	 				
\end{document}